\documentclass[sigconf,authorversion]{acmart}

\usepackage{acronym}
\usepackage{multirow}
\usepackage[inline]{enumitem}
\usepackage{array}

\usepackage{booktabs} 
\usepackage[skip=0pt]{caption}
\usepackage{pifont}

\usepackage[english]{babel}
\usepackage[utf8]{inputenc}
\usepackage[noend]{algpseudocode}
\usepackage{algorithmicx,algorithm}

\usepackage{CJKutf8}

\AtBeginDocument{%
  \providecommand\BibTeX{{%
    \normalfont B\kern-0.5em{\scshape i\kern-0.25em b}\kern-0.8em\TeX}}}

\AtBeginDocument{%
  \providecommand\BibTeX{{%
    \normalfont B\kern-0.5em{\scshape i\kern-0.25em b}\kern-0.8em\TeX}}}

\acrodef{US}{User Simulation}
\acrodef{USS}{User Satisfaction Simulation}

\renewcommand{\shortauthors}{Sun et al.}

\allowdisplaybreaks

\copyrightyear{2021}
\acmYear{2021}
\setcopyright{acmlicensed}\acmConference[SIGIR '21]{Proceedings of the 44th International ACM SIGIR Conference on Research and Development in Information Retrieval}{July 11--15, 2021}{Virtual Event, Canada}
\acmBooktitle{Proceedings of the 44th International ACM SIGIR Conference on Research and Development in Information Retrieval (SIGIR '21), July 11--15, 2021, Virtual Event, Canada}
\acmPrice{15.00}
\acmDOI{10.1145/3404835.3463241}
\acmISBN{978-1-4503-8037-9/21/07}



\author{Weiwei Sun$^1$*\qquad Shuo Zhang$^2$*\qquad Krisztian Balog$^3$\qquad Zhaochun Ren$^{1\dagger}$}
\author{Pengjie Ren$^1$\qquad Zhumin Chen$^1$\qquad Maarten de Rijke$^{4, 5}$}

\makeatletter
\def\authornotetext#1{
\if@ACM@anonymous\else
    \g@addto@macro\@authornotes{
    \stepcounter{footnote}\footnotetext{#1}}
\fi}
\makeatother
\authornotetext{Equal contribution.}
\authornotetext{Corresponding author.}

\affiliation{
 \institution{\textsuperscript{\rm 1}Shandong University, Qingdao \country{China}}
 \institution{\textsuperscript{\rm 2}Bloomberg, London \country{United Kingdom} \qquad \textsuperscript{\rm 3}University of Stavanger, Stavanger \country{Norway}}
 \institution{\textsuperscript{\rm 4}University of Amsterdam \qquad \textsuperscript{\rm 5}Ahold Delhaize Research, Amsterdam \country{The Netherlands}}
}
\email{sunnweiwei@gmail.com, szhang611@bloomberg.net,krisztian.balog@uis.no}
\email{{zhaochun.ren, chenzhumin}@sdu.edu.cn, jay.ren@outlook.com, m.derijke@uva.nl}

\def\authors{Weiwei Sun, Shuo Zhang, Krisztian Balog, Zhaochun Ren, Pengjie Ren, Zhumin Chen, and Maarten de Rijke}

\renewcommand{\shortauthors}{Sun et al.}

\begin{document}

\fancyhead{}

\title[Simulating User Satisfaction for the Evaluation of Task-oriented Dialogue Systems]{Simulating User Satisfaction for the Evaluation\\ of Task-oriented Dialogue Systems}


\begin{abstract}
Evaluation is crucial in the development process of task-oriented dialogue systems.
As an evaluation method, user simulation allows us to tackle issues such as scalability and cost-efficiency, making it a viable choice for large-scale automatic evaluation.
To help build a human-like user simulator that can measure the quality of a dialogue, we propose the following task: simulating user satisfaction for the evaluation of task-oriented dialogue systems.
The purpose of the task is to increase the evaluation power of user simulations and to make the simulation more human-like.
To overcome a lack of annotated data, we propose a user satisfaction annotation dataset, \ac{USS}, that includes 6,800 dialogues sampled from multiple domains,
spanning real-world e-commerce dialogues, task-oriented dialogues constructed through Wizard-of-Oz experiments, and movie recommendation dialogues. 
All user utterances in those dialogues, as well as the dialogues themselves, have been labeled based on a 5-level satisfaction scale.
We also share three baseline methods for user satisfaction prediction and action prediction tasks.
Experiments conducted on the \ac{USS} dataset suggest that distributed representations outperform feature-based methods. A model based on hierarchical GRUs achieves the best performance in in-domain user satisfaction prediction, while a BERT-based model has better cross-domain generalization ability.
\end{abstract}

\maketitle

\acresetall


\section{Introduction}

\begin{figure}[t]
 \centering
 \includegraphics[width=0.93\columnwidth]{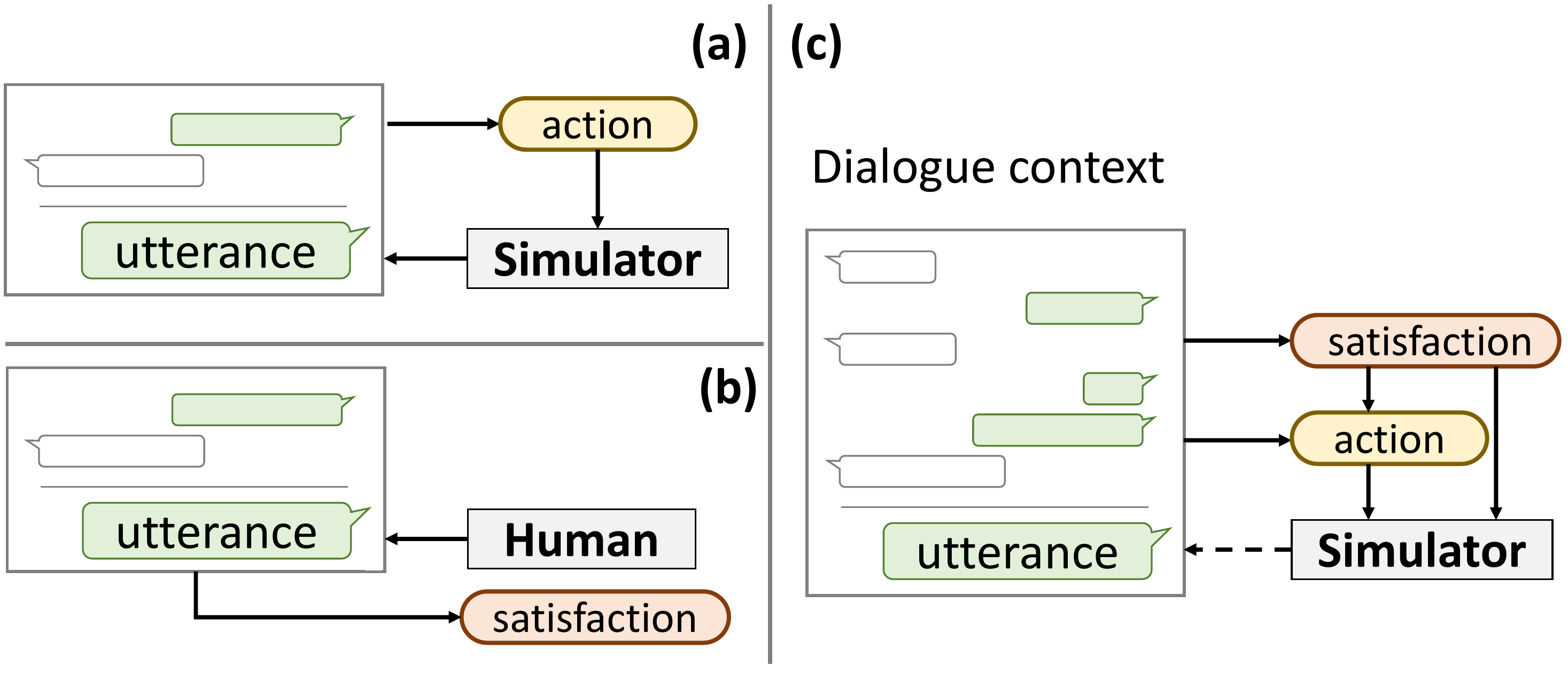}
 \vspace*{0.5\baselineskip}
 \caption{(a) Previous work on user simulation; (b) previous work on user satisfaction prediction; (c) our proposed task: simulating user satisfaction for evaluating task-oriented dialogues systems. We leave utterance generation (dotted line) as future work.}
 \label{figure:task}
\end{figure}

Task-oriented systems are developed to help users solve a specific task as efficiently as possible~\cite{Young2013POMDP}.
Evaluation is a crucial part of the development process of task-oriented dialogue systems.
For evaluating the performance of each module of a dialogue system, human evaluation, user satisfaction modeling, corpus-based approaches, and user simulation have all been leveraged~\citep{Deriu2020SurveyOE}.
Human evaluation through in-field experiments \citep{Lamel2000TheLA,Black2011SpokenDC} or crowd-sourcing \citep{Jurccek2011RealUE} is considered to reflect the overall performance of the system in a real-world scenario, but it is intrusive, time-intensive, and does not scale \citep{Deriu2020SurveyOE}.
User satisfaction modeling can be an alternative; it aims to automatically estimate user satisfaction based on human-machine interaction log data, but still requires human involvement.
To evaluate a dialogue system fully automatically, offline evaluation based on test sets is commonly used. However, this method is limited to a single turn and does not inform us about the overall usefulness of the system or about users' satisfaction with the flow of the dialogue~\citep{Zhang2020EvaluatingCR}. 
Therefore, evaluation results of offline methods have limited consistency with the results of human evaluation.
Simulation-based evaluation methods address the issues listed above; they are a viable choice for large-scale automatic evaluation~\citep{Deriu2020SurveyOE}.
User simulations can be used to evaluate functionalities of dialogue systems and they can serve as an environment to train reinforcement learning-based systems~\citep{Deriu2020SurveyOE}, leveraging agenda-based~\citep{Schatzmann2007AgendaBasedUS} or model-based simulation~\citep{Asri2016ASM}.
Building human-like user simulation is still an open challenge~\citep{Jannach2020ASO}.

\begin{figure}[t]
 \centering
 \includegraphics[width=0.95\columnwidth]{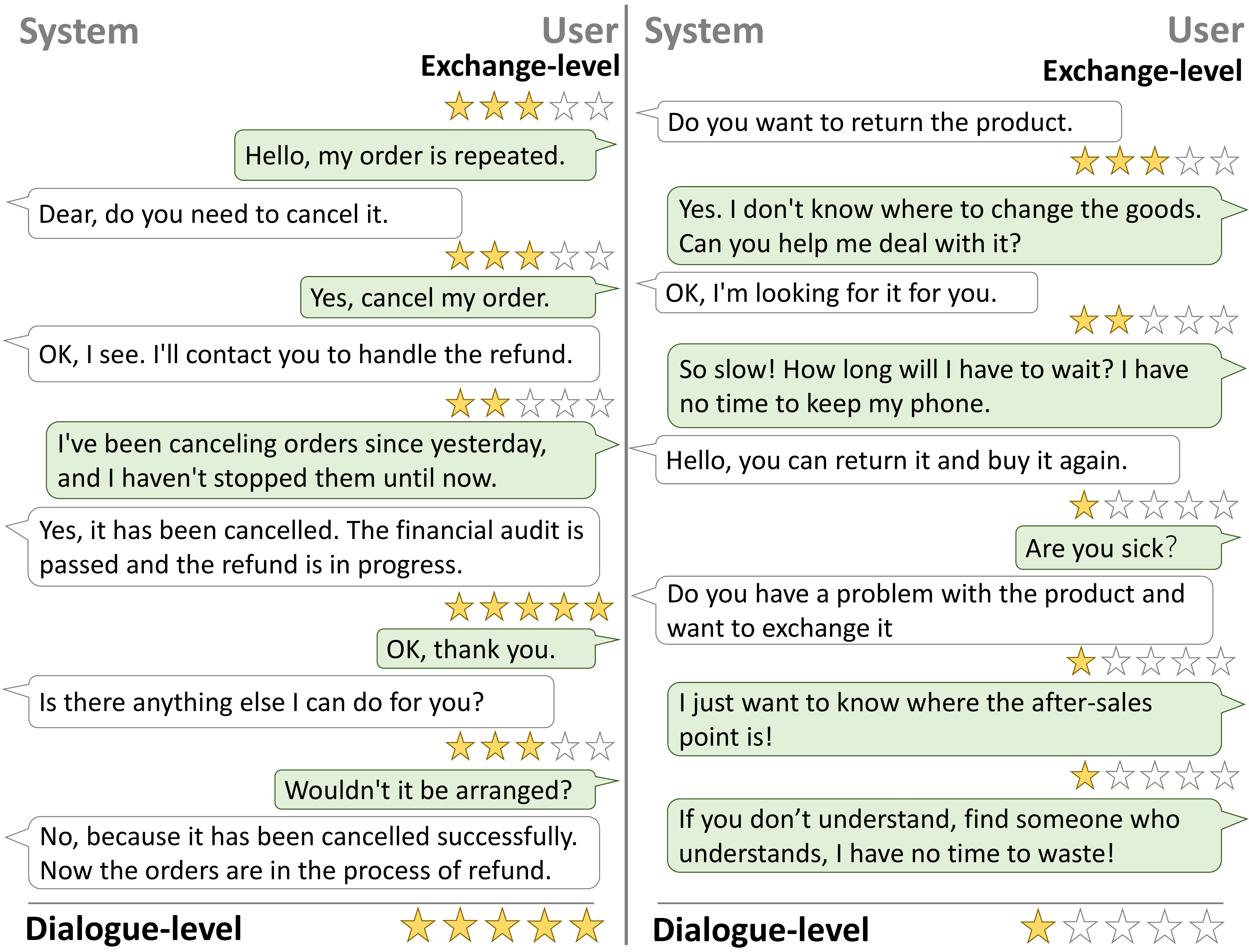}
 \vspace*{0.75\baselineskip}
 \caption{Two examples of dialogues in the JDDC dataset~\citep{Chen2020TheJC},
 with different degrees of user satisfaction. 
 The right-side system fails to understand the user's needs, and causes the user to be dissatisfied emotions and have a poor user experience. The left-side dialogue demonstrates an opposite case.}
 \label{figure:example}
 \vspace*{-0.25\baselineskip}
\end{figure}

To bridge the gap between human evaluation and user simulation, we attempt to combine user simulation with user satisfaction (cf.~Figure~\ref{figure:task}).
To this end, we first look into existing task-oriented dialogues and carry out a user study to investigate the characteristics of user satisfaction.
We arrive at two main observations:
\begin{enumerate*}[label=(\arabic*)]
\item \emph{User dissatisfaction is mainly caused by the system's failure in meeting the user's needs}.
Specifically, 36\% of the conversations are labeled as \emph{very dissatisfied}  because the system does not understand the user's needs, and 43\% are because the system understands the user's problems but cannot provide proper solutions. 
Figure~\ref{figure:example} illustrates the scenario.
\item \emph{Different degrees of satisfaction result in different sequences of user actions}.
For example, the right-side user in Figure~\ref{figure:example} may switch to customer service or explain further when encountering the same failed system reply in the context of different emotions.
We convert this intuition to a hypothesis that we verify by checking the records in the corpus.
When faced with a dialogue system's failure in understanding user needs, about 17.1\% of all users will switch to manual customer service, and about 64.3\% and 9.7\% will continue by providing additional information, or quit the conversation, respectively. This observation suggests that user simulation should work differently in different user satisfaction scenarios.
\end{enumerate*}

Informed by the observations just listed, we propose a novel task: \emph{to simulate user satisfaction for the evaluation of task-oriented dialogue systems}.
Figure~\ref{figure:task} illustrates the main difference between our task and previous work.
We extend the evaluation capability of user simulations and make the simulation more human-like by incorporating user satisfaction prediction and user action prediction. 

To facilitate research on user satisfaction simulation,
we develop a user satisfaction annotation dataset, \acfi{USS}. We invite 40 annotators to label both the dialogue level and exchange level user satisfaction of 5 commonly used task-oriented dialogue datasets in different domains. 
This results in a dataset of 6,800 dialogues, where each individual user utterance, as well as each complete dialogue, is labeled on a 5-point satisfaction scale. 
Each dialogue is labeled by 3 annotators; the expert ratings are highly correlated, with a Fleiss Kappa score of 0.574.
The \ac{USS} dataset shares some characteristics with existing datasets for user satisfaction, but also differs in important ways (see Table~\ref{table:dataset-comparision}):
\begin{enumerate*}
\item Our user satisfaction labeling occurs before the user utterance, and is based on the dialogue context between user and system instead of the satisfaction expressed in the user's utterance.
\item The \ac{USS} dataset includes multiple domains, such as e-commerce, reservations, recommendations, etc.
\item The \ac{USS} dataset exceeds existing user satisfaction data in scale.
\end{enumerate*}

We share three baseline approaches to perform satisfaction prediction and user action prediction based on the newly collected data in \ac{USS}: a feature-based method, a hierarchical GRU-based method, and a BERT-based method. 
Experimental results suggest that distributed representations outperform feature-based methods. 
The hierarchical GRU-based method achieves the best performance in in-domain user satisfaction prediction, while the BERT-based method has a better cross-domain generalization ability thanks to the pre-training. 
We also show that the BERT-based method achieves state-of-the-art performance on the action prediction task.

In summary, this paper makes the following contributions:
\begin{enumerate*}[label=(\arabic*)]
\item We propose the novel task of simulating user satisfaction for the evaluation of task-oriented dialogue systems.
\item We collect and share a dataset, \ac{USS}, that includes 6,800 annotated dialogues in multiple domains.
\item We introduce three baseline methods for the tasks of satisfaction prediction and action prediction using the \ac{USS} dataset.
\end{enumerate*}

\section{Related work}
Unlike chitchat systems, which focus on conversing with human on open domains, task-oriented dialogue systems aim to complete specific tasks for user~\cite{Wen2017ANE,lei2020interactive}.
Task-oriented dialogue systems can be divided into module-based and end-to-end-based methods~\cite{Jannach2020ASO}.
The former decomposes the dialogue system into four stages: language understanding, dialogue state tracking, dialogue policy learning, and response generation.
Recently, each stage in the module-based task-oriented dialogue systems has received increased attention~\cite{Hashemi2016QueryID,Yao2013RecurrentNN,Wen2017ANE,Mrksic2015MultidomainDS,Mrksic2017NeuralBT,Yan2017BuildingTD}.
End-to-end task-oriented dialogue systems rely on neural dialogue generation, which has received a lot of attention in recent years~\cite{Young2013POMDP,Banchs2013IRIS,Ameixa2014Luke}. 
Among all these approaches, sequence-to-sequence structure neural generation models~\cite{vinyals2015neural,li2016a,serban2016building,chen18www,lei2018sequicity,jin2018explicit} have been proved to be capable in multiple dialogue systems with promising performance. 

Evaluation is a crucial part of the development process of task-oriented dialogue systems.
Corpus-based approaches, user simulation, and user satisfaction modeling have all been leveraged~\citep{Zhang2020EvaluatingCR} for evaluating the performance of a task-oriented dialogue system.
Offline evaluation based on test sets is commonly used, but it is limited in scope to a single turn and does not inform us about the overall usefulness of the system or about users' satisfaction with the flow of the dialogue~\citep{Zhang2020EvaluatingCR}.
%
Employing simulation-based evaluation can tackle the above issues and become one viable choice for large-scale automatic evaluation~\citep{Deriu2020SurveyOE}.
User simulators are tools that are designed to simulate the user’s behavior, which can be used to train the dialogue manager in an offline environment \cite{Deriu2020SurveyOE} or to evaluate the dialogue policy \cite{Schatzmann2007AgendaBasedUS}. 
\citet{Eckert1997UserMF} propose the first statistical user simulator.
\citet{Scheffler2000ProbabilisticSO} propose a graph-based model.
\citet{Georgila2005LearningUS} use a Markov Model, and a hidden Markov model has been proposed by \citet{Cuayhuitl2005HumancomputerDS}.
\citet{Schatzmann2007AgendaBasedUS} propose an agenda-based user simulator, which represents the user state elegantly as a stack of necessary user actions, called the agenda.
\citet{Zhang2020EvaluatingCR} evaluate conversational recommender systems via an agenda-based user simulator.
Recent work employs neural approaches, esp.\ sequence-to-sequence models \cite{Asri2016ASM,Kreyssig2018NeuralUS}.
As far as we know, no previous study explicitly models the user satisfaction in user simulations.
Unlike previous work, we are the first to incorporate user satisfaction into user simulation to make the simulation more human-like.

Next to user simulations, user satisfaction modeling is the other evaluation method that is based on the idea that the usability of a system can be approximated by the satisfaction of its users \cite{Deriu2020SurveyOE}.
\citet{Ultes2013OnQR} note the impracticability of having a user rate a live dialogue. Thus, automatic prediction can be an alternative.
\citet{Walker1997PARADISEAF} propose the PARADISE framework, which estimates user ratings on the dialogue level. 
Evaluation methods that estimate user satisfaction at the exchange level have also been proposed \cite{Engelbrecht2009ModelingUS,Higashinaka2010IssuesIP,Hara2010EstimationMO}. 
They yield more fine-grained predictions and are especially useful for online dialogue breakdown detection. 
\citet{Schmitt2015InteractionQA} propose Interaction Quality (IQ) to assign user ratings by experts instead of real users.
\citet{Bodigutla2019MultidomainCQ} introduce the Response Quality (RQ) scheme to improve generalizability to multiple-domain conversations.

Unlike previous work on user satisfaction modeling, we simulate the user satisfaction changes without human involvement.
\begin{table}[]
\small
\centering
\setlength\tabcolsep{2pt}
\caption{Available datasets related to our task. AU/BU is short for After Utterance/Before Utterance.}
\label{table:dataset-comparision}
\begin{tabular}{l c l r r c c}
\toprule
\textbf{Dataset}      & \textbf{Year}    & \textbf{Domain} & \textbf{\#Dialog} & \textbf{\#Turns} & \textbf{Type} & \textbf{Level}
\\ 
\midrule
LEGO~\cite{Schmitt2012APA}     & 2012   & Bus  &  347   & 9,083 & AU  & 5
\\
IARD~\cite{Cai2020PredictingUI}   & 2020  & Movie  &  336     & 2,203 & AU & 2
\\
Alexa~\cite{Bodigutla2020JointTA}   & 2020  & Booking  & 3,129      & 20,167 & AU & 5
\\
MHCH~\cite{Liu2020TimeTT}   & 2020  & E-commerce  &  7,500    & 75,548 & BU  & 2
\\
\midrule
USS (Ours)   & 2021  & Multiple  &  6,800    &  99,569 & BU  & 5
\\
\bottomrule
\end{tabular}
\end{table}

\section{Task formulation}
\label{sec:tr}
To formulate the task of simulating user satisfaction, we first carry out a user study to explore the characteristics of user satisfaction in task-oriented dialogues.
Specifically, we invite 12 experts and let each expert annotate 20 dialogues sampled from the JDDC dataset; we used the JDDC dataset since it is more realistic than data constructed by the Wizard-of-Oz approach.
We ask each expert to score the user satisfaction for each dialogue turn and the entire conversation. In addition, a rational explanation is requested. 
We ask the experts to judge the user action changes after a change in satisfaction.
Based on this study, we answer the following questions:

\begin{enumerate*}[label=(\arabic*)]
\item \emph{What causes the user's dissatisfaction?}
%
%
We collect the results and find that, although annotators are satisfied with the system overall, about 12\% of the dialogue turns are labeled as unsatisfying. 
This indicates that there are fluctuations in user satisfaction when interacting with the system.
We analyze the annotators' explanations and find that the main reason for dissatisfaction relates to the system's failure to understand the user's needs or handling the user's requests.
Specifically, 36\% of all conversations labeled as \emph{very dissatisfied} are because \emph{the system does not understand the user's needs}, whereas 43\% are because \emph{the user does not approve the system's response}. In 64\% of the data, users had a bad user experience because \emph{the system was not professional enough or did not respond in time}.
Figure~\ref{figure:example} illustrates the scenario where the system does not understand the user's needs and causes low user satisfaction.

\item \emph{How does user satisfaction influence the user's behavior?}
Different degrees of satisfaction result in different sequences of user actions. 
Specifically, when encountering a failure in the a dialogue system's understanding of user needs, about 17.1\% of all users  \emph{switch to manual customer service}, and about 64.3\% and 9.7\% continue by \emph{providing additional information}, or \emph{quit the conversation}, respectively. 
Figure~\ref{figure:example} shows an example, where the right-side user switches to customer service or explains further when encountering the same failed system reply in light of different degrees of satisfaction.
Apart from user actions, we also observe changes such as attitude and information-seeking goal.
\end{enumerate*}

The above observations indicate that predicting the fluctuations of user satisfaction during interaction is non-trivial.
%
Thus, we formulate our research task, i.e., to \emph{simulate user satisfaction for the evaluation of task-oriented dialogue systems}. 

This simulation task focuses on the prediction of the next user action as well as user satisfaction. 
Suppose that we have a dataset $\mathcal{D} = \{(U_{i}, a_{i}, s_{i})\}_{i=1}^{N}$,  where for all $i \in [1,N]$, $U_{i}$ is the dialogue context, $a_{i}$ is the next-turn user action, and $s_{i}$ denotes user satisfaction.
The task objective is to learn a classification model $P(a, s \mid U)$ from $\mathcal{D}$, and thus given a dialogue context $U$, it predicts the next-turn user action $a$ and user satisfaction $s$ based on $P(a, s \mid U)$.
The purpose of the task to increase the evaluation power of user simulations and to make the simulation more human-like by incorporating the user's potential changes in satisfaction in a simulator.

\section{Constructing A Test Collection}
We propose a user satisfaction annotation dataset, \acfi{USS}. 
Below, we detail the creation of the dataset. We divide this section into $3$ phases: data preparation, user satisfaction assessment, and measures and disclaimers.

\subsection{Data preparation}

The \ac{USS} dataset is based on five benchmark task-oriented dialogue datasets: JDDC \cite{Chen2020TheJC}, Schema Guided Dialogue (SGD) \cite{Rastogi2020TowardsSM}, MultiWOZ 2.1 \cite{Eric2020MultiWOZ2A}, Recommendation Dialogues (ReDial) \cite{Li2018TowardsDC}, and
Coached Conversational Preference Elicitation (CCPE) \cite{Radlinski2019CoachedCP}. 
We first distinguish the user's emotion in the conversation by a classifier trained on annotated reddit data (weibo for Chinese), and then filter out all conversations that do not show negative emotions (i.e., anger, disgust, fear, sadness). 

\begin{enumerate*}[label=(\arabic*)]
\item JDDC is a large-scale, real-world Chinese e-commerce conversation corpus with over 1 million multi-turn dialogues. We first classify the conversation into 11 types according to the type of transaction, e.g., delivery, return, invoice, etc. Then, we sample 300 dialogue sessions from each type, for a total of 3,300 conversations. The JDDC data set provides the action of each user utterance, including 234 categories. We compress them into 12 categories based on a manually defined classification method.
\item SGD is a dataset consisting of over 20K annotated task-oriented conversations between a human and a virtual assistant spanning 16 domains. MultiWOZ 2.1 is a multi-domain dialogue dataset spanning 7 distinct domains and containing over 10K dialogues. We sample 1,000 conversations from the two datasets. We directly use the action annotation from the original datasets. The SGD has 12 actions, and MultiWOZ has 21 actions.
\item ReDial is an annotated dataset consisting of over 10K conversations, where users recommend movies to each other. We sample 1,000 dialogues. Since the original dataset does not provide actions, we use the action annotation provided by IARD \cite{Cai2020PredictingUI}.
\item CCPE is a dataset consisting of 502 dialogues with 12K annotated utterances between a user and an assistant discussing movie preferences. We sample 300 dialogues from the CCPE dataset and used the actions provided by the original dataset.
\end{enumerate*}

\subsection{User satisfaction assessment}
We hired 40 annotators to annotate exchange-level and dialogue-level user satisfaction levels of each conversation with five levels (1--5). 
We first show a dialogue between user and system in text form to the annotators and ask the annotators to label the user satisfaction of each user sentence at the \emph{exchange-level}. 
We require annotators to rate user satisfaction based on past conversations, so the satisfaction is assessed before the user’s sentence, not after writing the sentence. 
In this regard, we differ from previous annotation work \cite{Walker1997PARADISEAF,Schmitt2015InteractionQA,Bodigutla2019MultidomainCQ}.
The scale we asked annotators to follow was: 
\begin{enumerate*}[label=(\arabic*)]
\item Very dissatisfied (the system fails to understand and fulfill user’s request);
\item Dissatisfied (the system understands the request but fails to satisfy it in any way); 
\item Normal (the system understands users request and either partially satisfies the request or provides information on how the request can be fulfilled); 
\item Satisfied (the system understands and satisfies the user request, but provides more information than what the user requested or takes extra turns before meeting the request); and 
\item Very satisfied (the system understands and satisfies the user request completely and efficiently). 
\end{enumerate*}

Using a 5 point scale over a binary scale provides an option for the annotators to factor in their subjective interpretation of the extent of success or failure of a system’s response to satisfy a user’s request.
In addition, we ask the annotators to rate the \emph{dialogue-level} satisfaction to capture the overall satisfaction of a user’s interaction with the system.
We divide the data into two groups based on language, JDDC (Chinese) and Others (English). In each group, we randomly assign data to annotators to ensure that the different types of conversations in the group are evaluated according to a consistent standard. For the JDDC group, we also ask annotators to give a textual explanation for the rating.

\subsection{Measures and disclaimers}
To guarantee annotation quality, we ask at least three annotators to repeatedly label the data. If there is a discrepancy among the three annotators (i.e., three annotators give three different ratings), we ask a fourth annotator to recheck it. We removed the results of annotators that were inconsistent with others. Finally, expert ratings are highly correlated with a Fleiss Kappa score of 0.574. See Table~\ref{table:statistic} for descriptive statistics of the \ac{USS} dataset.

In all the provided instruction materials, we described the purpose of this data construction effort and pointed out that the data will only be used for research. 
We did not record any information about the annotators and warned the annotators not to divulge any of their private information. 

\begin{table}[t]
\small
\centering
\setlength\tabcolsep{2pt}
\caption{Statistics of the \ac{USS} dataset.}
\label{table:statistic}
\begin{tabular}{l rrrrr}
\toprule
\textbf{Domain}           & \textbf{JDDC}    & \textbf{SGD}  & \textbf{MultiWOZ}  & \textbf{ReDial} & \textbf{CCPE} \\ 
\hline
Language        & Chinese    &  English  & English & English & English \\ 
\#Dialogues        & 3,300      & 1,000    & 1,000 & 1,000 & 500 \\
Avg\# Turns & 32.3  &  26.7   &  23.1     &  22.5 & 24.9 \\ 
\hline
\#Utterances  & 54,517     &  13,833        &  12,553     & 11806      &  6,860 \\
~~Rating 1  & 120      &  5      & 12     & 20 & 10 \\
~~Rating 2  &  4,820  &  769    & 725    & 720 & 1,472\\
~~Rating 3  &  45,005  &  11,515  &  11,141 & 9,623 & 5,315\\
~~Rating 4   &  4,151  &  1,494   &  669   & 1,490 & 59\\
~~Rating 5  &  421      &  50    &  6     & 34 & 4\\
\bottomrule
\end{tabular}
\end{table}


\section{Experiments}

\subsection{Models used for comparison}
Inspired by previous work \cite{Jiao2019HiGRUHG,Yang2016HierarchicalAN,Barahona2021IsTU}, we consider three types of approach: Feature-based, RNN-based, and BERT. 

\subsubsection{Feature-based models}
We use (1) TF-IDF, (2) the length of the last utterance (i.e., the number of words), and (3) position of the current utterance as the features in feature-based models. 
We compare several machine learning models that have popularly been used for text classification \cite{Aggarwal2012MiningTD}: 
\begin{enumerate*}
\item logistic regression (LR), 
\item support vector machines (SVM), and
\item XGBoost.
\end{enumerate*}

\subsubsection{RNN-based models}
Given the dialogue context $U=\{u_{j}\}_{j=1}^{t}$, we first encode it to get the context representation $\mathbf{h}^{U}$, and then predict the user satisfaction by $P(s\mid U)=\operatorname{softmax}(\operatorname{MLP}(\mathbf{h}^{U}))$.
Inspired by previous work, we compare three methods for context representation encoding:
\begin{enumerate*}
\item GRU, which first concatenates the dialogue history into a long sentence, and then feeds the sentence into a Bidirectional GRU (BiGRU) model. Then the context representation is defined as the average pooled outputs of the BiGRU model.
\item HiGRU, which explores the hierarchical structure. First, it encodes each utterance in the dialogue using a word-level BiGRU to get the utterance representations $\mathbf{h}^{u_{j}}$.
Then it feeds the utterance representations into a sentence-level GRU, and define the context representation as the last hidden state of the sentence-level GRU~\cite{Jiao2019HiGRUHG}.
\item HiGRU+ATTN, which applies a two-level attention mechanism in HiGRU~\cite{Yang2016HierarchicalAN}. 
\end{enumerate*}

\subsubsection{BERT-based model}
Given the dialogue context $U=\{u_{j}\}_{j=1}^{t}$, we first concatenate it to a long sequence with $[SEP]$. Then we encode it into a latent representation via BERT~\citep{Devlin2019BERTPO}, and convert it into the condensed representation $\mathbf{h}^{U}$ through an average pooling operation. 
User satisfaction is predicted as $P(s\mid U)=\operatorname{softmax}(\operatorname{MLP}(\mathbf{h}^{U}))$.

\subsection{Implementation details}
To integrate the user satisfaction prediction and action prediction, we train two independent models for two tasks, in which action prediction takes the predicted output of satisfaction prediction model as the input. We use ground truth satisfaction in training and the model predicted satisfaction in testing. 
The Feature-based models are implemented using the scikit-learn toolkit.
For the BERT-based model, we use BERT-Base~(110M) pretrained weights\footnote{\url{https://github.com/huggingface/transformers}}~(hidden size is 768).
We use the BERT vocabulary~(size: 30,522) for all models (the Chinese BERT vocabulary for the JDDC domain), set the batch size $= 64$, the learning rate to 2e-5 for BERT and 1e-4 for others, use the AdamW optimizer~($\beta_1 = 0.9$, $\beta_2 = 0.999$, and $\epsilon$ = $10^{-8}$) to optimize parameters, use gradient clipping with a maximum gradient norm of 0.2, train up to 50 epochs on one NVIDIA TITAN RTX GPU, and select the best checkpoints based on performance on the validation set. Due to the serious imbalance of the satisfaction label, we up-sample the non-3 rating data during training. We take dialogue-level satisfaction as the last user utterance and use ``overall'' as the identification. As in previous work~\cite{Cai2020PredictingUI}, we use 10-fold cross-validation to evaluate the outcome.


\section{Evaluation}

\begin{table*}[!t]
\small
\centering
\setlength\tabcolsep{2pt}
\caption{Performance for user satisfaction prediction. Bold face indicates the best result in terms of the corresponding metric. Underline indicates comparable results to the best one.}
\label{table:satisfaction}
\begin{tabular}{l cccc cccc cccc cccc cccc}
\toprule
\multirow{2}{*}{\textbf{Domain}} 
& \multicolumn{4}{c}{\textbf{JDDC}} 
& \multicolumn{4}{c}{\textbf{SGD}}
& \multicolumn{4}{c}{\textbf{MultiWOZ}}
& \multicolumn{4}{c}{\textbf{ReDial}}
& \multicolumn{4}{c}{\textbf{CCPE}}
\\

\cmidrule(lr){2-5} \cmidrule(lr){6-9}  \cmidrule(lr){10-13}  \cmidrule(lr){14-17} \cmidrule(lr){18-21}

& UAR     & Kappa   & Rho  & F1
& UAR     & Kappa   & Rho  & F1 
& UAR     & Kappa   & Rho  & F1  
& UAR     & Kappa   & Rho  & F1  
& UAR     & Kappa   & Rho  & F1  
\\ 

\midrule

LR
& 0.221     &  0.054  & 0.400  & 0.011
& 0.211     &  0.049  & 0.251  & 0.005
& 0.214     &  0.042  & 0.599  & 0.009
& 0.211     &  0.040  & 0.240  & 0.008
& 0.214     &  0.060  & 0.669  & 0.025
\\
SVM
& 0.235     &  0.061  & 0.347  & 0.026
& 0.230     &  0.074  & 0.169  & 0.020
& 0.215     &  0.030  & 0.425  & 0.021
& 0.209     &  0.038  & 0.205  & 0.015
& 0.212     &  0.027  & 0.534  & 0.040
\\
XGBoost
& 0.205     & 0.007   & \textbf{0.584}  & 0.003
& 0.202     & 0.011   & 0.442  & 0.001
& 0.200     & 0.002   & 0.690  & 0.001
& 0.207     & 0.030   & 0.391  & 0.002
& 0.200     & 0.001   & 0.707  & 0.004
\\
\midrule
HiGRU+ATTN
& 0.330     & 0.115   & 0.502  & \underline{0.180}
& 0.262     & 0.082   & \underline{0.475}  & 0.058
& 0.224     & \underline{0.142}   & 0.842  & 0.197
& \textbf{0.261}     & 0.097   & \textbf{0.441}  & 0.118
& 0.223     & 0.109   & 0.869  & 0.214 
\\
HiGRU
& \textbf{0.339}     & 0.126   & 0.524  & 0.171
& \textbf{0.293}     & \textbf{0.118}   & 0.451  & \textbf{0.086}
& 0.225     & \textbf{0.143}   & \textbf{0.886}  & \textbf{0.238}
& \underline{0.257}     & 0.084   & 0.324  & 0.083
& \textbf{0.237}     & \textbf{0.167}   & 0.881  & \textbf{0.274}
\\
GRU
& 0.302     & 0.092   & 0.497  & 0.132
& 0.245     & 0.072   & 0.248  & 0.027
& 0.231     & 0.105   & 0.813  & 0.167
& 0.254     & 0.104   & 0.421  & 0.121
& 0.226     & 0.124   & 0.880  & 0.207 
\\
\midrule
BERT
& 0.329     & \textbf{0.131}  & 0.554  & \textbf{0.185}
& 0.261     & 0.094   & \textbf{0.477}  & 0.048
& \textbf{0.256}     & 0.133   & 0.823  & 0.224
& \underline{0.257}     & \textbf{0.122}   & 0.390  & \textbf{0.125}
& \underline{0.232}     & 0.147   & \textbf{0.891}  & 0.245
\\
\bottomrule
\end{tabular}

\end{table*}


\begin{table*}[!t]
\small
\centering
\setlength\tabcolsep{2pt}
\caption{Performance for user action prediction. Bold face indicates the best result in terms of the corresponding metric. Underline indicates comparable results to the best one.}
\label{table:action}
\begin{tabular}{l cccc cccc cccc cccc cccc}
\toprule
\multirow{2}{*}{\textbf{Domain}} 
& \multicolumn{4}{c}{\textbf{JDDC}} 
& \multicolumn{4}{c}{\textbf{SGD}} 
& \multicolumn{4}{c}{\textbf{MultiWOZ}}
& \multicolumn{4}{c}{\textbf{ReDial}}
& \multicolumn{4}{c}{\textbf{CCPE}}
\\

\cmidrule(lr){2-5} \cmidrule(lr){6-9}  \cmidrule(lr){10-13}  \cmidrule(lr){14-17} \cmidrule(lr){18-21}

& Acc     & Prec   & Recall  & F1
& Acc     & Prec   & Recall  & F1
& Acc     & Prec   & Recall  & F1
& Acc     & Prec   & Recall  & F1
& Acc     & Prec   & Recall  & F1
\\ 

\midrule

LR
&  0.565    & 0.208   & 0.123  & 0.133
&  0.460    & 0.321   & 0.308  & 0.309
&  0.414    & 0.150   & 0.130  & 0.134
&  0.495    & 0.467   & 0.472  & 0.464
&  0.509    & 0.325   & 0.314  & 0.316
\\
SVM
& 0.493     & 0.214   & 0.139  & 0.147
& 0.451     & 0.344   & 0.351  & 0.345
& 0.374     & 0.141   & 0.138  & 0.135
& 0.459     & 0.423   & 0.444  & 0.427
& 0.462     & 0.327   & 0.327  & 0.322  
\\
XGBoost
& 0.621     & 0.270   & 0.138  & 0.165
& 0.516     & 0.395   & 0.370  & 0.370
& 0.479     & 0.226   & 0.126  & 0.139
& 0.593     & 0.540   & 0.509  & 0.506
& 0.553     & 0.380   & 0.349  & 0.356
\\
\midrule
HiGRU+ATTN
& \textbf{0.623}     & 0.363   & 0.176  & 0.194
& 0.617     & 0.498   & 0.481  & 0.481
& 0.487     & 0.221   & 0.152  & 0.155
& 0.590     & 0.548   & 0.512  & 0.488
& 0.611     & 0.421   & 0.408  & 0.411
\\
HiGRU
& 0.618     &  0.370  & \underline{0.196}  & \textbf{0.229}
& 0.643     &  0.534  & 0.505  & 0.507
& \underline{0.518}     &  0.216  & 0.162  & 0.167
& \textbf{0.622}     &  \textbf{0.584}  & \textbf{0.532}  & \textbf{0.534}
& \underline{0.672}     &  0.503  & 0.472  & 0.482 
\\
GRU
& 0.598     &  0.337  & 0.166  & 0.187
& 0.444     &  0.322  & 0.304  & 0.298
& 0.460     &  0.211  & 0.124  & 0.129
& 0.599     &  0.536  & 0.494  & 0.457
& 0.545     &  0.550  & 0.354  & 0.354
\\
\midrule
BERT
& 0.614     & \textbf{0.391}   & \textbf{0.199}  & \underline{0.224}
& \textbf{0.661}     &  \textbf{0.570}  & \textbf{0.572}  & \textbf{0.560}
& \textbf{0.519}     &  \textbf{0.255}  & \textbf{0.183}  & \textbf{0.191}
& 0.614     &  0.573  & \underline{0.531}  & \underline{0.530}
& \textbf{0.674}     &  \textbf{0.696}  & \textbf{0.495}  & \textbf{0.496}
\\
\bottomrule
\end{tabular}

\end{table*}


\begin{table}[]
\small
\centering
\setlength\tabcolsep{2pt}
\caption{Cross-domain performance for user satisfaction prediction. Report UAR.}
\label{table:cross-domain}
\begin{tabular}{ll cccc}
\toprule
\textbf{From} & \textbf{To}
& \textbf{SGD}  & \textbf{MWOZ} & \textbf{ReDial}  & \textbf{CCPE}
\\

\midrule
\multirow{3}{*}{\textbf{SGD}}
& SVM
& \textcolor{gray}{0.230}     & 0.209   & 0.211  & 0.198
\\
& HiGRU
& \textcolor{gray}{0.293}     & 0.240   & 0.230  & 0.212
\\
& BERT
& \textcolor{gray}{0.261}     & \textbf{0.249}   & \textbf{0.254}  & 0.223
\\
\midrule
\multirow{3}{*}{\textbf{MWOZ}}
& SVM
& 0.208     & \textcolor{gray}{0.215}   & 0.206  & 0.208
\\
& HiGRU
& 0.224     & \textcolor{gray}{0.225}   & 0.221  & 0.219
\\
& BERT
& \textbf{0.233}     & \textcolor{gray}{0.256}   & 0.219  & 0.226
\\
\midrule
\multirow{3}{*}{\textbf{ReDial}}
& SVM
& 0.216     & 0.227   & \textcolor{gray}{0.221}  & 0.199
\\
& HiGRU
& 0.211     & 0.221   & \textcolor{gray}{0.261}  & 0.220
\\
& BERT
& 0.228     & 0.218   & \textcolor{gray}{0.257}  & \textbf{0.239}
\\
\midrule
\multirow{3}{*}{\textbf{CCPE}}
& SVM
& 0.217     &  0.208   & 0.218  & \textcolor{gray}{0.214}
\\
& HiGRU
& 0.211     & 0.223   & 0.227  & \textcolor{gray}{0.237}
\\
& BERT
& 0.216     & 0.213   & 0.219  & \textcolor{gray}{0.232}
\\
\bottomrule
\end{tabular}

\end{table}


\subsection{Evaluation metrics}
For the user satisfaction prediction task, following \citep{Schmitt2015InteractionQA}, we use the \emph{Unweighted Average Recall} (UAR), the arithmetic average of all class-wise recalls, a linearly weighted version of \emph{Cohen’s Kappa}, and \emph{Spearman’s Rho} as evaluation metrics. 
We also use the \emph{F1-score} for the \emph{dissatisfactory} (rating $<3$) class as the binary classification metric, as most turns and dialogues belong to the \emph{satisfactory} (rating $\geq 3$) class.
For the user action prediction task, we use
\emph{Accuracy} (Acc, the proportion of predicted correct labels over the total number of predicted and actual labels for every utterance), 
\emph{Precision} (Prec, the proportion of the predicted correct labels over the number of predicted labels), 
\emph{Recall} (the proportion of the predicted correct labels over the number of actual labels), and 
the \emph{F1-score} (the harmonic mean of precision and recall) as evaluation measures.

\subsection{Experimental results}

Table~\ref{table:satisfaction} shows the results for the user satisfaction prediction task.
The best results in terms of the corresponding metric are shown in bold. If there are multiple similar best results, we show them all underlined.
In general, HiGRU achieves the best overall performance (e.g., an absolute improvement of +3 for UAR, +2 for Kappa, and +4 for F1 over BERT in SGD data). BERT and HiGRU+ATTN can achieve performance comparable to HiGRU, followed by GRU.
Among the 3 feature-based methods, SVM performs best, followed by LR. XGBoost is significantly weaker than other methods in all metrics, except Rho.
Table~\ref{table:satisfaction} further shows that all deep learning methods perform better than feature-based metrics.

Table~\ref{table:action} shows the results for the user action prediction task.
In general, the BERT-based model performs best among all methods, followed by HiGRU.
BERT outperforms HiGRU on all performance measures except for the ReDial data, possibly due to the lack of sufficient training data.
Among the 3 feature-based methods, XGBoost achieves the best performance, obtaining an absolute improvement of about +6 for Acc, +7 for Prec, +3 for Recall, and +4 for F1 compared to LR. XGBoost also outperforms GRU in many metrics.

\subsection{Analysis}
Since we have multiple domains in the dataset, we further analyze the cross-domain generalization capabilities of the user satisfaction prediction model. 
Table~\ref{table:cross-domain} shows the results. The rows and columns in Table~\ref{table:cross-domain} indicate training data and test data, respectively (e.g., 0.233 in the first column of the sixth row indicates that a BERT model trained on MultiWOZ can get a UAR score of 0.233 on SGD data). 
In terms of datasets, the models trained on SGD and MultiWOZ get the best performance on each other’s data respectively, and the models trained on ReDial get the best performance on CCPE, possibly due to the similarity between domains. The model trained on CCPE has relatively poor generalization ability, possibly due to limited training data size.
In terms of methods, BERT achieves better generalization performance than SVM and HiGRU, possibly due to the improvement of pre-training on the large-scale corpus.


\section{Utilization of this Resource}

We have developed resources that are meant to help answer the question of what is a good dialogue.
Our annotations and prediction task offer a better characterization of what is a good dialogue than existing datasets.
Exchange-level user satisfaction and action prediction can reflect what kind of system behavior will bring positive user satisfaction and what behavior will harm the user experience, which makes our method applicable to many related fields. 

\subsection{Building human-like user simulation}
In most prior work, user simulations mechanically give the slots, and thus measure very limited aspects of a dialogue. 
Building a human-like user simulation remains an open challenge. In this study, we propose the task of user satisfaction simulation and release a dataset for the task. Inspired by previous work on similar tasks \citep{Jiao2019HiGRUHG,Yang2016HierarchicalAN,Barahona2021IsTU}, we provide a series of baselines. 
However, due to the challenging nature of the task, there is plenty of room to improve user satisfaction prediction, and to explore how user satisfaction prediction can be combined with action prediction. 
Response generation based on user satisfaction (i.e., reflect user satisfaction in a generated utterance) is still an open problem. Previous work on open-domain dialogue may serve as a reference \citep{Zhou2018EmotionalCM}.
In addition to user satisfaction, how to ground a user simulator by introducing external knowledge~\cite{sun2021conversations,Meng2020DukeNetAD,xu2020conversational,ma2020compare} and persona~\cite{Li2016APN} to establish a more human-like user simulator has not yet been studied.

\subsection{Future applications}
The \ac{USS} dataset can be used not only for user simulation but also for other conversational information access tasks. As a user satisfaction annotation dataset that exceeds existing ones in scale, our data can facilitate research on user satisfaction modeling~\cite{Pragst2016RecurrentNN} and POMDP-based dialogue systems~\cite{Lemon2012DataDrivenMF,Young2013POMDP}.
Moreover, the \ac{USS} dataset can also facilitate research into dialogue breakdown detection, and human-machine hand-off prediction~\cite{Liu2020TimeTT}.
In the JDDC domain, we provide annotators' explanations on user satisfaction annotations, which includes a total of 9,900 explanation texts.
This information can be applied to user studies of user satisfaction, and interpretability studies of evaluations.


\section{Conclusion}
We have proposed the task of simulating user satisfaction for evaluating task-oriented dialogue systems, so as to enhance the evaluation of dialogue systems. 
We have collected and released a new benchmark dataset, namely \ac{USS}, for the proposed task. 
Our dataset contains a total of 6,800 dialogues spanning multiple domains. We have introduced three baselines for our task: feature-based, RNN-based, and BERT-based methods. 
%
Experiments conducted on the newly collected dataset suggest that distributed representations do outperform feature-based methods. Besides, HiGRU achieves the best performance in in-domain user satisfaction prediction, while a BERT-based method has better cross-domain generalization ability.

As to our future work, we would like to continue to investigate the combination of the user satisfaction prediction and action prediction task, and response generation based on user satisfaction. 


\section*{Data}
We share the \ac{USS} dataset at \url{https://github.com/sunnweiwei/user-satisfaction-simulation}.
\begin{acks}
This work was supported by the National Key R\&D Program of China with grant No. 2020YFB1406704, the Natural Science Foundation of China (61972234, 61902219, 62072279),
the Key Scientific and Technological Innovation Program of Shandong Province (2019JZZY010129),
the Tencent WeChat Rhino-Bird Focused Research Program (JR-WXG-2021411), the Fundamental Research Funds of Shandong University, the Hybrid Intelligence Center,
a 10-year program funded by the Dutch Ministry of Education, Culture and Science through
the Netherlands Organisation for Scientific Research, \url{https://hybrid-intelligence-centre.nl}. 
All content represents the opinion of the authors, which is not necessarily shared or endorsed by their respective employers and/or sponsors.
\end{acks}

\newpage
\bibliographystyle{ACM-Reference-Format}
\bibliography{references}

\newpage

\end{document}